\pdfoutput=1

\documentclass[fleqn,usenatbib,usedcolumn,letters]{mnras}
\usepackage{natbib}

   \usepackage[british]{babel}             
   \usepackage{newtxtext}                  
   \usepackage[slantedGreek]{newtxmath}    
  
   \let\la=\lesssim 
  
   \usepackage[T1]{fontenc}               
   \usepackage{graphicx}

\usepackage{graphicx}	
\usepackage{amsmath}	
\usepackage{amssymb}	

\title[Chemical evolution of the Gaia Sausage]{The Fall of a Giant. Chemical evolution of Enceladus, alias the Gaia Sausage}

\author[Vincenzo et al.]
{Fiorenzo Vincenzo$^{1}$\thanks{E-mail: f.vincenzo@bham.ac.uk}, Emanuele Spitoni$^{2}$, Francesco Calura$^{3}$, 
Francesca Matteucci$^{4,5,6}$, 
\newauthor 
Victor Silva Aguirre$^{2}$,  Andrea Miglio$^{1}$, Gabriele Cescutti$^{5}$
\\ ~ \\
$^{1}$School of Physics and Astronomy, University of Birmingham, Edgbaston,  B15 2TT, UK \\
$^{2}$Stellar Astrophysics Centre, Department of Physics and Astronomy, Aarhus University, Ny Munkegade 120, DK-8000 Aarhus C, Denmark \\
$^{3}$INAF, Osservatorio Astronomico di Bologna, Via Gobetti 93/3, 40129 Bologna, Italy\\
$^{4}$Dipartimento di Fisica, Sezione di Astronomia, Universita` di Trieste, via G.B. Tiepolo 11, I-34131, Trieste, Italy \\ 
$^{5}$INAF, Osservatorio Astronomico di Trieste, via G.B. Tiepolo 11, I-34131, Trieste, Italy \\ 
$^{6}$INFN, Sezione di Trieste, Via Valerio 2, I-34100 Trieste
}

\date{Accepted 2019 May 7. Received 2019 May 1; in original form 2019 March 8}

\begin{document}

\label{firstpage}

\pagerange{\pageref{firstpage}--\pageref{lastpage}} \pubyear{2019}

\maketitle

\begin{abstract}
We present the first chemical evolution model for Enceladus, alias the Gaia Sausage, to investigate the star formation history of one of the most massive satellites  accreted by the Milky Way during a major merger event. Our best chemical evolution model for Enceladus nicely fits the observed stellar [$\alpha$/Fe]-[Fe/H] chemical abundance trends, and reproduces the observed stellar metallicity distribution function, by assuming low star formation efficiency, fast infall time scale, and mild outflow intensity. We predict a median age for Enceladus stars $12.33^{+0.92}_{-1.36}$ Gyr, and -- at the time of the merger with our Galaxy ($\approx10$ Gyr ago from Helmi et al.) -- we predict for Enceladus a total stellar mass $M_{\star} \approx 5 \times 10^{9}\,\text{M}_{\sun}$. By looking at the predictions of our best model, we discuss that merger events between the Galaxy and systems like Enceladus may have inhibited the gas accretion onto the Galaxy disc at high redshifts, heating up the gas in the halo. This scenario could explain the extended period of quenching in the star formation activity of our Galaxy about $10$ Gyr ago, which is predicted by Milky Way chemical evolution models, in order to reproduce the observed bimodality in [$\alpha$/Fe]-[Fe/H] between thick- and thin-disc stars. 
\end{abstract}




\section{Introduction} 

Understanding the past mass assembly history and the dynamical evolution of the stellar components of our Galaxy by looking at its stellar halo is one of the major challenges of contemporary astrophysics (e.g. \citealt{morrison2000,prantzos2008,deason2017,helmi2018,dimatteo2018,hayes2018}). 
During its evolution across cosmic times, it is highly likely that the Milky Way (MW) was surrounded by many galaxy companions, which suffered from strong tidal interactions, being continuously stripped of their stars and gas by the gravitational pulling forces; a large fraction of these companions are today seen as stellar streams, or dispersed tidal debris in the velocity- and chemical-abundance spaces \citep{helmi1999a,helmi2018,kruijssen2018,simion2019,gallart2019}.

From a theoretical point of view, the standard $\Lambda$-cold dark matter (CDM) paradigm for the formation and evolution of the structures in the cosmos predicts a large number of satellite galaxies around massive disc galaxies like the Milky Way. In particular, the MW dark matter (DM) halo should have formed from the accretion of filamentary structures and from the coalescence of many small DM halos at high-redshifts, with the mass of the accreted systems increasing -- on average -- as a function of time \citep{helmi1999,bullock2001}. 

The absence of a large number of dwarf galaxies gravitationally bound to the MW was one of the major discrepancies between observations and model predictions \citep{klyptin1999}, until the advent of the Sloan Digital Sky Survey (SDSS) and the Dark Energy Suvey (DES), which -- in the last fifteen years -- discovered an impressive amount of new dSphs and ultra-faint dwarfs (UfDs) around the Milky Way \citep{simon2019}.

Thanks to the SDSS and -- more recently -- to the DES and SDSS-Gaia, many stellar streams were discovered in the inner MW halo as faint substructures  \citep{belokurov2006,shipp2018,myeong2018}. The most studied of these stellar streams is associated to Sagittarius dSph, and was intensively investigated in terms of its chemical abundances, kinematics, and stellar population properties by many independent studies over the years  \citep{ibata2001,majewski2003,newberg2003,deboer2015}. Other stellar streams were later the subject of similar intense investigations \citep{koposov2017,deboer2018a,deboer2018b,koposov2019}.

Recent unprecedented analysis of the seven-dimensional Gaia-SDSS catalogue revealed a metal-rich component in the inner Galaxy halo, which shows a peculiar elongated shape along the horizontal axis of the velocity ellipsoid, as given by the azimuthal stellar velocity component, $v_{\theta}$, versus the radial velocity component, $v_{r}$ \citep{belokurov2018}. This renamed ``sausage'' in the velocity-space is probably due to relatively metal-rich stars compared to the Galaxy halo (with metallicities $Z \approx Z_{\sun}/10$), which have also large velocity anisotropy ($\beta \approx 0.95$) (see \citealt{myeong2018b,fattahi2019}). This ``sausage'' represents the dynamical record in the velocity-space of a head-on major collision that the MW experienced more than $10\,\text{Gyr}$ ago with a quite massive dwarf galaxy. We also address the readers to the works of \citet{iorio2019}, for a detailed study of the dynamical structure of the MW halo by making use of RR Lyrae, and \citet{dimatteo2018,haywood2018,mackereth2019}, for interesting studies on the connection between the Gaia Sausage and the MW accretion history from chemical and kinematical points of view, using different techniques. The progenitor (now disrupted) galaxy of this ``sausage'' in the velocity-space is now called Gaia Enceladus, or Gaia Sausage. 

 A sample of confirmed Gaia Sausage member stars are present in the catalogue of the APO Galactic Evolution Experiment (APOGEE). These stars were selected by \citet{helmi2018}, to show that -- in the $[\alpha/\text{Fe}]$-$[\text{Fe}/\text{H}]$ chemical abundance diagram -- they have properties very similar to those of some dSph stars, i.e.  tipically lower $[\alpha/\text{Fe}]$ values than metal-poor MW halo stars (see, also, \citealt{hayes2018} for a detailed study). 
 
 Interestingly, \citet{nissen2010} showed that, with  VLT/UVES and NOT/FIES observations,  the metal-rich tail of  the Galactic halo is characterized by  two distinct populations of stars. The authors explained such a dichotomy by proposing that the low-[$\alpha$/Fe] stellar component was probably accreted from dwarf galaxies. Finally, before the release of Gaia DR2, \citet{fernandez2018} investigated the average star formation rate (SFR) and initial mass function (IMF) in a very similar sample of APOGEE stars with respect to that later selected by \citet{helmi2018} for Enceladus, finding two distinct [$\alpha$/Fe]-sequences. 

  In this Letter, we present the first attempt of modelling in detail the chemical evolution of Enceladus, fitting our chemical evolution model to reproduce the observed $[\alpha/\text{Fe}]$-$[\text{Fe}/\text{H}]$ and the metallicity distribution function (MDF) of the stars in Enceladus. This Letter is structured as follows. In Section \ref{sec2} we describe the adopted chemical evolution model and the main features of the best model for Enceladus. In Section \ref{sec3} we present the results of our study. Finally, in Section \ref{sec4}, we draw our conclusions.


\section{The chemical evolution model} \label{sec2}

We develop a chemical evolution model to reproduce the abundances in Enceladus, by assuming the same set of stellar nucleosynthetic yields as in \citet{francois2004} for massive stars (dying as core-collapse SNe), Type Ia SNe, and asymptotic giant branch stars. For Type Ia SNe, we assume the single degenerate scenario, with the same prescriptions as in \citet{matteucci2001}. In our model, we solve a set of differential equations, and assume stellar nucleosynthetic yields, which are the same as those of the MW two-infall chemical evolution model of \citet{spitoni2018}. 
The stellar yields and Type Ia SN model are usually selected with the aim of reproducing the observed MW chemical abundances, and later 
applied to study also external galaxies. We follow this approach, relying on the nucleosynthesis assumptions of \citet{spitoni2018}. 

We assume that Enceladus forms at high redshift from the rapid collapse of primordial gas in the intergalactic medium (IGM). The infall rate of gas from the IGM into the Enceladus potential well is of the form $\mathcal{I}(t) = A\,e^{-t/\tau_{\text{inf}}}$, where $\tau_{\text{inf}}$ is the infall time-scale, a free parameter of the model, and the normalisation constant, $A$, fixes the total amount of gas mass accreted from the IGM during the galaxy lifetime, which is the so-called infall mass, $M_{\text{inf}}$. 
The star formation rate (SFR) in our model follows a linear Schmidt-Kennicutt law, i.e. $\text{SFR}=\text{SFE}\times M_{\text{gas}}$, with $\text{SFE}$ being the star formation efficiency. We also assume galactic winds to develop when the thermal energy of the gas -- heated by stellar winds and SNe -- exceeds the binding energy of the gas due to the galaxy potential well, as in \citet{bradamante1998}. The intensity of the outflow rate is directly proportional to the SFR, namely $\mathcal{O}(t)=w\times\text{SFR}(t)$, where $w$ is the mass loading factor, a free parameter of the model. In order to compute the binding energy, we assume that the mass of the DM halo is $M_{\text{DM}}=10\times M_{\text{inf}}$. 
Finally, in our model we assume the stellar lifetimes of \citet{padovani1993} and the initial mass function (IMF) of \citet{kroupa1993}, defined in the mass range between $0.1$ and $100\,\text{M}_{\sun}$. 

We create a grid of $\sim17,500$ models by varying the main free parameters ($\mathrm{SFE}$, $w$, and $\tau_{\mathrm{infall}}$), in order to reproduce the trend of the observed $[\alpha/\text{Fe}]$-$[\text{Fe/H}]$ \citep{helmi2018}. In our best model, we assume $\text{SFE}=0.42\,\text{Gyr}^{-1}$, $\tau_{\text{inf}}=0.24\,\text{Gyr}$, and the infall law is normalised to have a total amount of accreted gas mass $M_{\text{infall}} = 10^{10}\,\text{M}_{\sun}$. The mass loading factor does not play an important role in our models for Enceladus, because the wind is predicted to occur relatively late in the galaxy evolution, at [Fe/H] abundances larger than those observed for Enceladus stars by \citet{helmi2018}; we assume in our fiducial model $w=0.5$, but in Section \ref{sec3} we will show our results for $w=0$, $0.5$, and $1.0$. 

\subsection{Exploring the parameter space} 

\begin{figure}
	\centering
	\includegraphics[width=7.0cm]{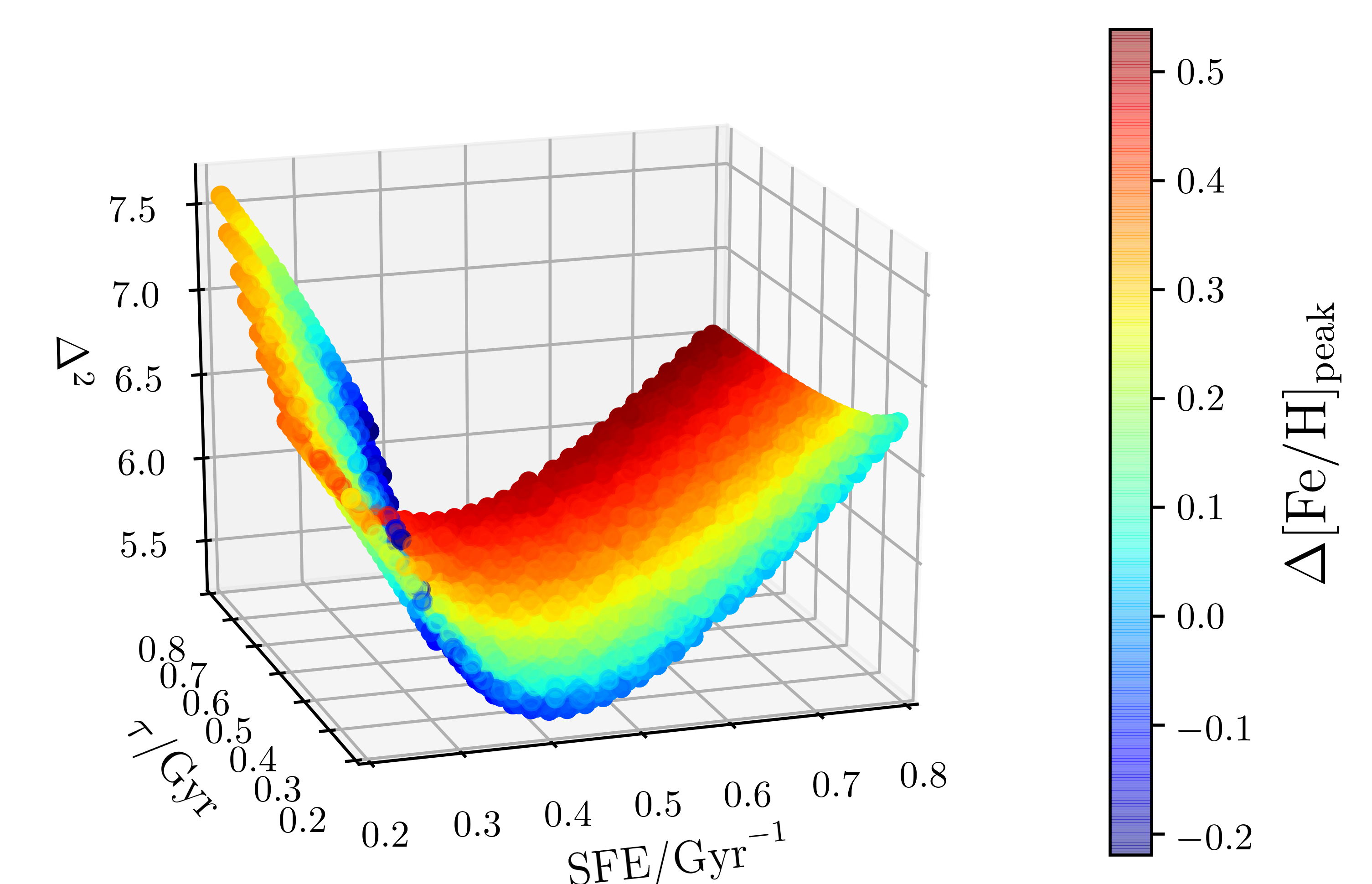} 
    \caption{  The adopted figure-of-merit to evaluate the goodness of our models in reproducing 
    the observed [$\alpha$/Fe]-[Fe/H] of Enceladus stars ($\Delta^{2} \propto \sum_{i}{\Big[ [ \alpha/\text{Fe}](\text{[Fe/H]}_{i})_{ \text{mod} }^{2} - [\alpha/\text{Fe}](\text{[Fe/H]}_{i})_{ \text{obs} }^{2} \Big]}/\sigma^{2}_{i}$) 
    as a function of SFE and $\tau_{\text{inf}}$. The colour coding represents the difference 
    between the [Fe/H] of the MDF peak of each model and that of the best model, which reproduces both the observed stellar [$\alpha$/Fe]-[Fe/H] and the observed 
    MDF (the absolute minimum at $\text{SFE}=0.42\,\text{Gyr}^{-1}$ and $\tau_{\inf}=0.24\,\text{Gyr}$). }
    \label{fig0}
\end{figure}

On the one hand, our predictions for $[\alpha/\text{Fe}]$-$[\text{Fe}/\text{H}]$ and the MDF are highly sensitive to the SFE, which determines the time (and [Fe/H] value) when the first Type Ia SNe explode, causing $[\alpha/\text{Fe}]$ to sharply decrease as a function of [Fe/H]. On the other hand, the infall time scale cannot be precisely constrained if we only look at $[\alpha/\text{Fe}]$-$[\text{Fe}/\text{H}]$. Nevertheless, the MDF --  which is an important observational constraint to reproduce -- strongly depends on the assumed infall time scale; this is shown in Fig. \ref{fig0}, where our estimator to evaluate the goodness of the models to reproduce [$\alpha$/Fe]--[Fe/H] is drawn as a function of SFE and $\tau_{\text{inf}}$. The best model, reproducing both the observed [$\alpha$/Fe]--[Fe/H] and the observed MDF, corresponds to the minimum at SFE$=0.42\,\text{Gyr}^{-1}$ and $\tau_{\inf}=0.24\,\text{Gyr}$.


\section{Results}  \label{sec3}

\begin{figure}
	\centering
	\includegraphics[width=6.5cm]{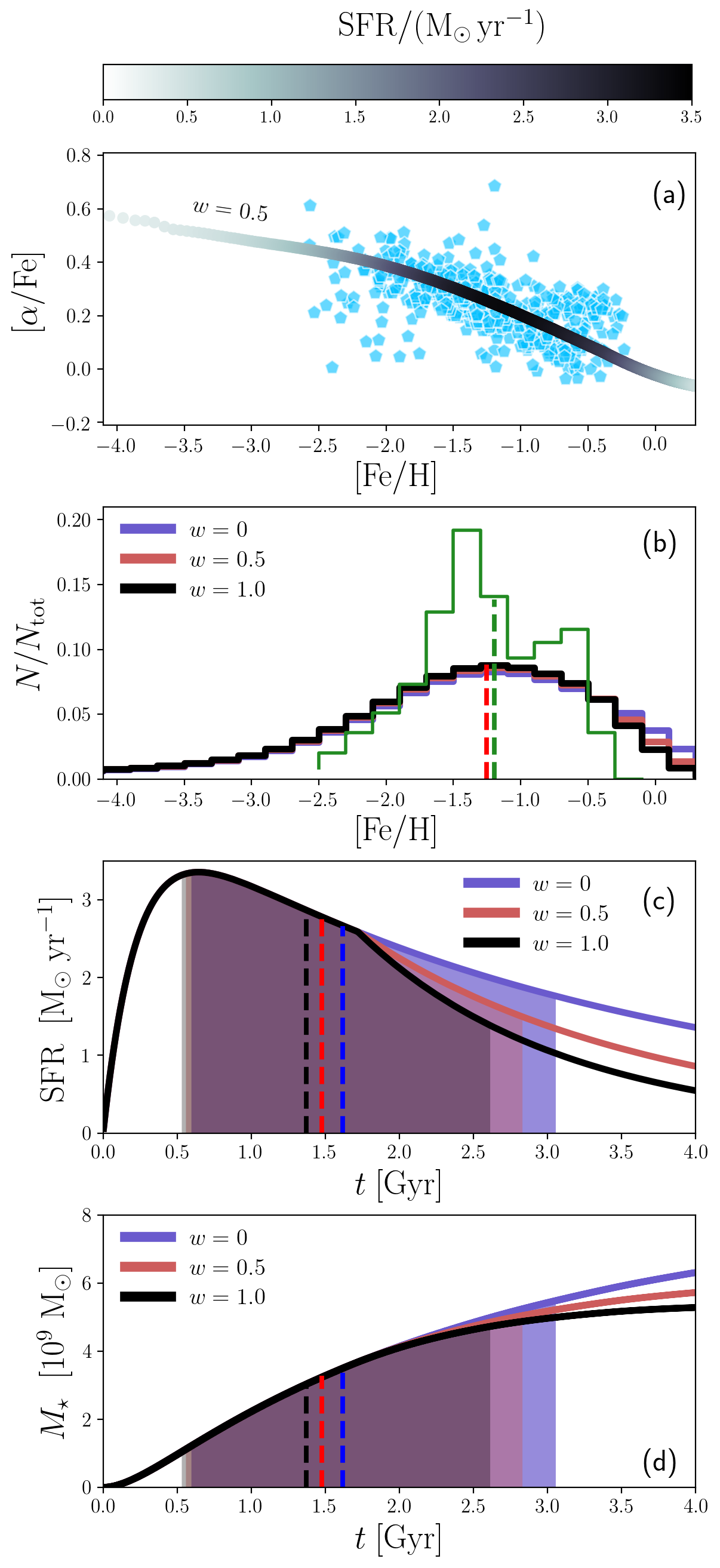} 
    \caption{  \textit{(a)} Observed [$\alpha$/Fe]-[Fe/H] for Enceladus from \citet{helmi2018} (light blue pentagons), compared with our fiducial chemical evolution model with $w=0.5$. The color code indicates the intensity of the SFR. \textit{(b)} The MDF of our fiducial model for Enceladus, by varying the mass loading factor, $w$. The green histogram corresponds to the data of \citet{helmi2018}. The vertical dashed lines indicate the median [Fe/H] abundance of our best model with $w=0.5$ (in red) and data (in green). \textit{(c)} The time evolution of the SFR predicted by our models for Enceladus with different values of $w$. Each vertical line labels the median stellar birth time of the corresponding model, and the shaded area indicates the $\pm$1 $\sigma$ region. \textit{(d)} The time evolution of the stellar mass predicted by our models with different $w$ for Enceladus.  }
    \label{fig1}
\end{figure}

In Fig. \ref{fig1}(a) we compare our fiducial chemical evolution model for Gaia Enceladus with the observed data of \citet{helmi2018}. The colour coding in the figure corresponds to the predicted SFR along the chemical evolution track as the system evolves as a function of time. Given the relatively low SFE of the best model, we can reproduce the overall declining trend of Enceladus stars in the $[\alpha/\text{Fe}]$-$[\text{Fe}/\text{H}]$ abundance diagram, which is due to large contribution of Fe from Type Ia SNe, contributing already at such low $[\text{Fe/H}]$. Our model can reproduce the observed trend in the \citet{helmi2018} data set. Our model indicates that, when most of the observed stars formed, Enceladus was characterised by SFR values between $\sim1$ and $3.5\,\text{M}_{\sun}\,\text{yr}^{-1}$.

In Fig. \ref{fig1}(b) we compare the MDF of our fiducial chemical evolution model for Enceladus 
with the data set of \citet{helmi2018}. To show that the mass loading factor plays a marginal role for the bulk of the chemical evolution of Enceladus, 
we show models with different values of $w$, from $w=0$ (no wind) to $w=1.0$. 
We find that the median iron abundance of our fiducial model with $w=0.5$ is $[\text{Fe/H}]=-1.26^{+0.82}_{-1.06}$ dex,
in very good agreement with the observed median value $[\text{Fe/H}]_{\text{obs}}=-1.21^{+0.59}_{-0.49}$ dex. 
The predicted MDFs show a large spread in the iron abundances, while the data of \citet{helmi2018} do not cover the very low metallicity regime. 
However, it is worth to stress that it is very difficult to obtain very accurate spectroscopic chemical abundance measurements for very metal-poor stars (see the discussion in \citealt{placco2018}). 
Secondly, it is plausible that a non-negligible fraction of the oldest, most metal poor stars formed in Enceladus did not reach the inner Galactic halo; 
the oldest stellar components of galaxies typically have, in fact, the highest velocity dispersion \citep{sanders2018,ting2018,mackereth2019b}.

From the analysis of the best model, we can reconstruct the total stellar and gas mass of Enceladus before the collision with the MW 
(which happened $\sim10$ Gyr ago, according to \citealt{helmi2018}), 
as well as the ages of Enceladus stars. 
In Fig. \ref{fig1}(c), we show our predictions for the evolution of the SFR as a function of time, for our fiducial model, for different mass loading factors. 
By assuming for the age of the Universe $t_{\text{U}}=13.8$ Gyr, and considering only the SFH from $t=0$ to $4$ Gyr, when the merger with the Galaxy 
approximately took place \citep{helmi2018}, the median age of Enceladus stars from our model with $w=0.5$ 
corresponds to $t_{\text{med}}=12.33^{+0.92}_{-1.36}$ Gyr. 
Assuming different mass loading factors does not significantly change our predictions for the 
median ages of the stars in Enceladus (see the different vertical lines in Fig. \ref{fig1}c); in particular, increasing the mass loading factor 
determines a slight increase also in the median stellar ages of the model. 

\begin{figure}
	\centering
	\includegraphics[width=6.5cm]{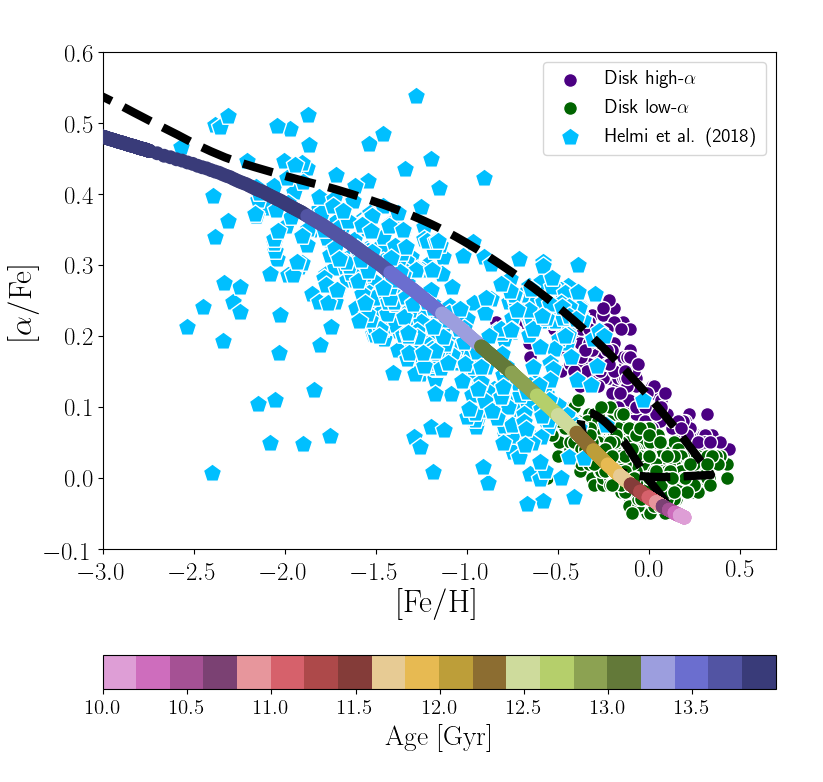}
    \caption{Observed [$\alpha$/Fe]-[Fe/H] abundance diagram from \citet{helmi2018} for Enceladus stars (light blue filled pentagons)  along with the MW disc data from the APOKASC sample \citep[high-$\alpha$ and low $\alpha$ sequence in purple and green points, respectively]{silvaaguirre2018}. The dashed black line represents the two-infall chemical evolution model of \citet{spitoni2018}, whereas the solid colour-coded line is our fiducial model for Enceladus, with the colour code indicating the predicted ages of the stars.}
    \label{fig2}
\end{figure}

Our predicted SFH in Fig. \ref{fig1}(c) globally accounts for the ages derived by \citet{helmi2018}, which are in the range $10$-$13$ Gyr, and our best 
chemical evolution model is remarkably in agreement with \citet{gallart2019}, 
which find a median age of Enceladus stars $t_{\text{med}}=12.37$ Gyr from isochrone-fitting, using a very 
different approach. Moreover, our results are also in agreement with the short SFR time scales inferred by \citet{fernandez2018} for their APOGEE sample 
of stars. 
Finally, in Fig. \ref{fig1}(d), we show the predicted evolution of the total stellar mass of Enceladus. 
In particular, the stellar and gas masses of Enceladus at $t_{\text{med}}$ as predicted by our model with $w=0.5$ 
are $M_{\star,\text{Enc}} = 3.24 \times 10^9 \,\text{M}_{\sun}$ and $M_{\text{gas},\text{Enc}} = 6.62 \times 10^9\,\text{M}_{\sun}$, respectively, with 
gas fraction $\approx0.67$. At the time of the merger with the Milky Way, 
therefore, we predict a stellar mass $M_{\star}\approx5\times10^{9}\,\text{M}_{\sun}$, in agreement with \citet{belokurov2018,helmi2018}. 


In Fig. \ref{fig2}, we compare the observed data of \citet{helmi2018} for [$\alpha$/Fe]-[Fe/H] in Enceladus with the observed MW abundances of 
thick and thin-disc stars 
from the APOKASC sample \citep{silvaaguirre2018}. 
In the figure, we also compare the predicted [$\alpha$/Fe]-[Fe/H] chemical evolution track of the two-infall model 
of \citet{spitoni2018} with our best model for Enceladus assuming $w=0.5$, 
where the colour coding represents the predicted age of the stars in Enceladus. 
The model of \citet{spitoni2018} is remarkable in the context of the two-infall chemical evolution models for the MW, because it has been 
developed to reproduce at the same time the observed age distribution of MW thick- and thin-disc stars from asteroseismic analysis of Kepler light curves 
(providing the most precise method to determine the stellar ages) and the 
chemical abundances from APOGEE,  for a sample of stars in common between APOGEE and Kepler \citep{silvaaguirre2018}. 


In order to reproduce the observed bimodal distribution of MW thick- and thin-disc stars in [$\alpha$/Fe]-[Fe/H], together with 
the distribution of the stellar ages, the model of \citet{spitoni2018} revised the classical two-infall model for the MW of \citet{chiappini1997}, by assuming 
a second infall which started $\approx 9.4$ Gyr ago, with a time delay of $\sim4.5$ Gyr after the beginning of the first infall. 
Such a time delay between the two infall episodes is much longer than that assumed by all previous 
two-infall chemical evolution models for the MW \citep{chiappini1997,grisoni2017,Sahijpal2018}, 
but it agrees with other recent independent studies, 
like \citet{noguchi2018} or \citet{grand2018}, which obtained very similar findings, 
but without attempting to fit also the observed age distribution of MW disc stars, as done in \citet{spitoni2018}. 
In summary, recent studies of the MW chemical evolution seem to agree that an extended hiatus in the Galactic SFH at high redshifts 
is required to reproduce the observed bimodality in the [$\alpha$/Fe]-[Fe/H] between MW thick- and thin-disc stars. 

By comparing the predicted age distribution of the stars in our best chemical evolution model for Enceladus and the predictions of the two-infall chemical evolution model 
of \citet{spitoni2018} for the MW, we propose that the mechanism -- which quenched the MW star formation at high redshifts, heating up the gas in the DM halo -- 
was a major merger event with a satellite like Enceladus (see, for example, \citealt{gabor2010,pontzen2017,vandevoort2018,hunt2018} for the quenching mechanisms in galaxies, and \citealt{dimatteo2008,martin2017,wilson2019} for the effects of mergers on high-redshift galaxies). 

We note that there might be a time-sequence problem between the age of the merger indicated by \citet{helmi2018} ($\approx10$ Gyr ago) and the findings of the two-infall model  for the MW of \citet{spitoni2018}, where the second infall happened $\approx9.4$ Gyr ago (see their figure 2). 
However, a $\sim10$ per cent error in the ages of old stellar populations is well within the uncertainties of the most precise methods currently available for determining ages of field stars 
(e.g., asteroseismology; see also \citealt{silvaaguirre2018}), and this can thereby conciliate the merger time from \citet{helmi2018} with the \citet{spitoni2018} model for the MW.  



\section{Discussion and conclusions} \label{sec4}

We have presented the first attempt of modelling the chemical evolution of Enceladus, in order to reproduce the $[\alpha/\text{Fe}]$-$[\text{Fe}/\text{H}]$ abundance trend and the MDF 
observed by \citet{helmi2018}. Our fiducial model assumes  $\text{SFE}=0.42\,\text{Gyr}^{-1}$, mass loading factors $w=0.5$, 
infall time scale $\tau_{\text{infall}}=0.24\,\text{Gyr}$, and infall mass $M_{\text{infall}} = 10^{10}\,\text{M}_{\sun}$. 
Our main findings and conclusions can be summarised as follows. 

\begin{enumerate}
    \item We find for Enceladus a median iron abundance  $\text{[Fe/H]} = -1.26^{+0.82}_{-1.06}$, obtained from the predicted galaxy SFH. Our result is in agreement with observations, which suggest a median iron abundance $[\text{Fe/H}]_{\text{obs}} = -1.21^{+0.59}_{-0.49}$ \citep{helmi2018}.     
    
    \item According to our results, the median age of Enceladus stars is $12.33^{+0.92}_{-1.36}$ Gyr, remarkably in agreement with \citet{gallart2019}, which estimated a median age 
    $\approx12.37$ Gyr from isochrone-fitting analysis, by using a very different approach with respect to ours. We note that the position in the CMD of the stars with $\text{[Fe/H]=-1.3}$ (and $[\alpha/\text{Fe}]=0.22$) in the sample of \citet{helmi2018} could be reproduced with isochrones of $\sim 13$ Gyr. 
    
            \item The predicted age distribution of the stars from our best Enceladus chemical evolution model corroborates the time of the merger, occurring about $10$ Gyr ago, estimated by \citet{helmi2018}, because the large majority of the stars in our best model have ages larger than $10$ Gyr. 
    
    \item  We predict that the stellar mass of Enceladus at the epoch of the merger suggested by \citet{helmi2018} is $M_{\star} \approx 5 \times 10^{9}\,\text{M}_{\sun}$, in agreement with the findings of \citet{belokurov2018,helmi2018,mackereth2019},  with a gas fraction $\approx 0.67$ at the median age of Enceladus stars. The predicted Enceladus stellar mass is comparable with the predicted MW stellar mass from \citet{spitoni2018} at the same epoch, which is $M_{\star,\text{MW}} \approx 8 \times 10^{9}\,\text{M}_{\sun}$. Since we assume in our chemical evolution model for Enceladus an infall mass $M_{\text{infall}}=10^{10}\,\text{M}_{\sun}$, it is unlikely that Enceladus alone provided sufficient gas mass to assemble the thin disc of our Galaxy.

    \item The merger between Enceladus and our Galaxy was likely the cause of a temporary quenching of the star formation and gas accretion in our Galaxy at high redshifts, which can be seen also in the predicted SFH of the two-infall model of \citet[see their figure 4, upper panel]{spitoni2018} but also in the predicted bimodal SFH of the best MW-like galaxy in the cosmological hydrodynamical simulation of \citet{grand2018}, which is characterised by an extended quenching phase, occurring approximately at the same epochs of the merger indicated by \citet{helmi2018}, before the formation of the Milky Way thin disc. This temporal sequence is corroborated by our best model for Enceladus.
    
    \item In the context of the two-infall chemical evolution model for the MW, it is likely that Enceladus was cannibalised by the Galaxy towards the end of the first infall episode, as a part of the gas and sub-structures of the infall episode itself. Nevertheless, before the collision between the MW and Enceladus happened, there might have been strong tidal interactions between the two galaxies as well, and this likely influenced the early accretion of gas from the IGM onto the Galaxy disc.

    \item Further investigations are needed to confirm with different techniques and with higher precision the age distribution of the stars in Enceladus. Asteroseismology techniques currently provide the best way to probe stellar interiors, to determine with very high accuracy stellar ages \citep{casagrande2016,silvaaguirre2016,miglio2017}. In the future, asteroseismology combined with chemodynamical simulations will allow us to study the mass assembly history of our Galaxy with unprecedented temporal resolution, and this will be the subject of our future work.  
    
\end{enumerate}

Finally we note that an optimal chemical element to test different theories of halo formation may be barium \citep{spitoni2016}, which is (relatively) easily measured in low-metallicity stars (see, for more details, \citealt{cescutti2006} and subsequent papers of the same author). In particular, \citet{spitoni2016} demonstrated that the predicted [Ba/Fe]-[Fe/H] relation in dSphs and UfDs is quite different than that in the Galactic halo; it will be interesting, in the future, to investigate the abundance trends of neutron-capture s-process elements in Enceladus stars, to be compared with similar abundance trends in the MW galaxy.


\section*{Acknowledgements}
\begin{small}
We thank an anonymous referee for their comments, which greatly improved the clarity of this work. We thank 
T. Mackereth, J. Montalban, G. Iorio, and P. E. Nissen for interesting and stimulating discussions. 
FV and AM acknowledge support from the European Research Council Consolidator Grant funding scheme (project ASTEROCHRONOMETRY, 
G.A. n. 772293). ES and VSA acknowledge support from the Independent Research Fund Denmark (Research grant 7027-00096B). 
Funding for the Stellar Astrophysics Centre is provided by The Danish National Research Foundation (Grant agreement no.: DNRF106). 
FC acknowledges funding from the INAF PRIN-SKA 2017 program 1.05.01.88.04. FM acknowledges research funds from the University of Trieste (FRA2016). 
\end{small}












\bsp	
\label{lastpage}
\end{document}